\documentclass{emulateapj}
  \usepackage{graphicx}
  \usepackage{epstopdf}
  \usepackage{natbib}

\usepackage{amssymb}
\usepackage{natbib}
\usepackage{amsmath}

\newcommand{\ltsima}{$\; \buildrel < \over \sim \;$}
\newcommand{\ltsim}{\lower.5ex\hbox{\ltsima}}

\newcommand{\be}{\begin{equation}}
\newcommand{\ee}{\end{equation}}
\newcommand{\bea}{\begin{eqnarray}}
\newcommand{\eea}{\end{eqnarray}}

\newcommand{\phz}{photo-\emph{z}~}

\begin{document}
\DeclareGraphicsExtensions{.pdf,.png,.jpg,.gif}

\title{Characterizing and Propagating Modeling Uncertainties in
Photometrically-derived Redshift Distributions}

\author{Augusta Abrahamse\altaffilmark{1,2}, Lloyd Knox\altaffilmark{1}, Samuel Schmidt\altaffilmark{1}, Paul Thorman\altaffilmark{1}, J.~Anthony Tyson\altaffilmark{1}, Hu Zhan\altaffilmark{3}}
\altaffiltext{1}{Department of Physics,
 1 Shields Avenue,
 University of California,
 Davis, CA 95616, USA.}
\altaffiltext{2}{Universidad Privada Boliviana, Av. Capitan Victor Ustariz Km. 6.5,
Cochabamba, Bolivia}
\altaffiltext{3}{National Astronomical Observatories,Chinese Academy of Sciences,
 A20 Datun Rd, Chaoyang District,
 Beijing 100012, China }

\begin{abstract}
The uncertainty in the redshift distributions of galaxies has a significant potential impact on the cosmological parameter values inferred from multi-band imaging surveys.  The accuracy of the photometric redshifts measured in these surveys depends not only on the quality of the flux data, but also on a number of modeling assumptions that enter into both the training set and spectral energy distribution (SED) fitting methods of photometric redshift estimation. In this work we focus on the latter, considering two types of modeling uncertainties: uncertainties in the SED template set and uncertainties in the magnitude and type priors used in a Bayesian photometric redshift estimation method. We find that SED template selection effects dominate over magnitude prior errors.  We introduce a method for parameterizing the resulting ignorance of the redshift distributions, and for propagating these uncertainties to uncertainties in cosmological parameters.
\end{abstract}
\maketitle

\section{Introduction \label{sec:intro}}

The surprising discovery of the accelerated expansion of the universe
\citep{riess98,perlmutter99} has invigorated efforts to determine the history of the expansion rate to high precision.  By providing a greater understanding of the cosmic acceleration, further observational input holds exciting prospects for improving our understanding of the fundamental laws of physics as well as the fate and possibly the origin of the universe \citep{DETF2006}.  Many observational efforts are aimed at determining the distance-redshift relation via a variety of distance-dependent signals including the apparent magnitude of supernovae, the angular location of the acoustic feature in the galaxy number density correlation function, and the amplitudes of tomographic cosmic shear power spectra. See \citet{Bassett:2009mm, Howell:2009mt,  Heavens:2008ud, Zhan:2008jh, Huterer:2010hw} for recent reviews.

Determining the redshifts associated with these signals can be challenging. In many cases, due to the large numbers and the faintness of the objects being used for distance determinations, we must extract redshift information not from spectroscopy, but from photometry in multiple broad bands \citep{baum62,connolly95}.  Redshifts determined from such data are called photometric redshifts, or simply photo-\emph{z}'s.  Recent work has emphasized the stringent requirements on the quality of redshift information needed to avoid significantly degrading or biasing inferences about the dark energy posited to explain cosmic acceleration \citep{FernandezSoto:2001xp, Huterer:2004tr, Ma:2005rc,  DETF2006, Zhan:2006gi, Abdalla:2007uc, Bridle:2007ft, Bernstein:2007uu}.

Traditional tomographic analysis of photometric data for cosmic shear and galaxy clustering begins with a sorting of galaxies into photo-$z$ bins.  This binning is a useful
stage of data compression which, if done correctly, leads to very
little loss of information.  These bins can be small in number and
hence fairly coarse.  As \citet{Ma:2005rc} have shown (in the
context of a low-dimensional dark energy model and a relatively shallow survey) cosmological parameter
uncertainties do not decrease significantly in going from five to more
photo-$z$ bins.  However, significant information loss can occur if the redshift probability distribution for a given galaxy, $P(z)$, is discarded immediately after binning.

The analysis procedure we present here also begins with a sorting of galaxies.  We refer to this sorting process as `collecting' rather than `photo-$z$ binning' to emphasize two points:  1) the sorting criterion, (or possibly criteria) is not restricted to redshift estimates and 2) even if the sorting is based on a redshift estimate, the distribution of the collection of galaxies with redshift will extend beyond the photo-$z$ redshift range used to define the collection.

However the collection is defined, a weak lensing map and a galaxy
count map can then be made for each collection.  From these maps, all the auto and
cross-power spectra can then be calculated \citep{hu99}, as well as other summary
statistics such as bispectra \citep{Takada:04a} and shear peak counts \citep{Wang:09a}.

The crucial redshift information that we need in order to calculate
model predictions for  these power spectra (and any other summary statistics)
is the redshift distribution, $dN/dz$, for each collection \citep{Ma:2005rc, Huterer:2005ez, Zhan:2006gi}.  In this paper, as discussed in \citet{Cunha:2009} and \citet{Wittman:2009sv}, we sum individual $P(z)$'s for the galaxies in each collection in order to infer the $dN/dz$ for that collection.

Our estimates of $dN/dz$ are {\em model-dependent}.  They depend on
assumptions about the distribution of galaxies with respect to spectral
type, apparent magnitude in a reference band and redshift.
We analytically demonstrate that, if the modeling assumptions are correct
that go into our $dN/dz$ estimates, then these estimates are unbiased (Appendix~\ref{sec:bias}).

Of course the modeling assumptions cannot be perfectly correct.  Thus
a crucial element of our methodology is a treatment of the uncertainty in $dN/dz$
that arises from uncertainty in the modeling.  Our treatment allows for this uncertainty
to be propagated through to the rest of the datwea reduction process.  The method for
determining the uncertainty in $dN/dz$ in a way that can be fully propagated to scientific
conclusions is the chief contribution of this paper.

Past work on propagation of redshift errors through to the cosmological parameters has typically relied on highly idealized models for the distribution of the redshift errors.  The redshift distributions considered are often far simpler than the distributions expected for real photometric surveys (Ilbert, et al. 2009).  For instance \citet{Ma:2005rc} assume the distribution of \phz estimate for given true redshift, $P(z_{ph}|z_{true})$, to be Gaussian, parameterized by a mean and scatter as a function of redshift. \citet{Huterer:2005ez} assume that photo-$z$ errors take the form of a bias that varies with $z$.  In reality, degeneracies in color space for multiband photometric determination of redshifts can result in redshift error distributions that are asymmetrical and multi-peaked, with peaks separated by $\Delta z > 1$.  As demonstrated by a growing body of research, it is important to take into account these catastrophic \phz errors and to add complexity to the \phz uncertainty models \citep{Schneider:2006br, Amara:2007as, Bernstein:2007uu, Sun:2008ge, Bernstein:2009bq, Nishizawa:2010vr, Hearin:2010jr}.

Although sufficiently robust spectroscopic sampling may reduce the need for modeling assumptions and the resulting uncertainties, for proposed large-scale future weak lensing surveys such as LSST, PanSTARRS and DES, there is much to be gained from using very large numbers of faint and distant galaxies for the distance determinations. In the case of LSST, several billion galaxies with $i$ band magnitudes as faint as 25 will be used.  It is very  challenging to spectroscopically determine redshifts for a fair sample of such faint objects. Spectroscopic samples preferentially contain galaxies with clear spectral features and are therefore biased towards specific galaxy types. Moreover they are significantly volume incomplete several magnitudes brighter than the faint limits of the photometry.  In other words, training sets used to develop photometric model inputs may be unrepresentative of the larger galaxy population, resulting in uncontrolled errors in the \phz estimation. Thus, a method for quantifying such modeling uncertainties is an absolute necessity for understanding the resulting uncertainties in the redshift distributions and the subsequent uncertainties in the cosmological parameters.


We present here a method for parameterizing the redshift error distributions that is completely general and thus accommodates the existence of catastrophic errors. The parameterization arises naturally out of the consideration of specific modeling assumptions and data. For specificity we use the LSST data model (described in \S\,\ref{sec:templatenoise}) when giving examples of sensitivity to model assumptions.

We show that changes in modeling assumptions result in variations in the redshift distribution, $dN/dz$, which can be captured by principal component analysis (PCA), where just a few modes are sufficient to represent these effects. The $dN/dz$ can be reproduced from linear combinations of these modes with the weights on each mode becoming the parameters of the new representation. The uncertainty on each weight is extracted from the variations resulting from uncertainties in modeling assumptions, and this information is then usable for any likelihood analysis into which \phz uncertainties enter. These new variables are continuous and lend themselves to a Fisher matrix or MCMC analysis of the parameter space.

In \S\,\ref{sec:zinference} we review Bayesian (model-based) methods for inferring redshifts from multi-band photometry, and we introduce principal component analysis (PCA) as a method for naturally capturing the modeling uncertainties in photometric redshift estimation. In \S\,\ref{sec:templatenoise} we present a model demonstrating how uncertainty in the selection of a set of spectral energy distribution (SED) templates can be captured by PCA. In \S\,\ref{sec:magpriors} we explore uncertainties in the type and magnitude distributions that can be used as priors on photometric redshift estimates \citep{Benitez:1998br} and demonstrate again the utility of PCA in parameterization of the modeling uncertainty. In \S\,\ref{sec:otherwork} we relate our results to work that has propagated \phz uncertainties through to constraints on dark energy.
In \S\,\ref{sec:conclusions} we summarize our findings.

\section{Inferring redshifts from multi-band photometry}
\label{sec:zinference}

Our method takes as a starting point a calculation of the probability distribution of the redshift of an object given its measured colors and certain modeling assumptions.  We then present a simple estimator for the redshift distributions, $dN/dz$, of collections of such galaxies.  For us, the ``collections'' will consist of galaxies all in the same photometric redshift bin.  Of course, our modeling assumptions may be incorrect.  Thus we present a method for parameterizing the resulting uncertainty in the $dN/dz$ for each photometric redshift bin.  With the uncertainty in the $dN/dz$ parameterized, and the uncertainty in the parameters specified, the problem is now well-posed for further propagation of the uncertainty in the photo-\emph{z}'s to uncertainty in the cosmological parameters.  The following subsections lay out each of these steps.

\subsection{Bayesian Calculation of Redshift Probability Distributions}

Let us consider $t$ to be a variable (or set of variables) that uniquely specifies the SED of a galaxy. Then specifying for a galaxy its $t$, redshift $z$, and apparent magnitude in a single reference wavelength band, $m_1$, is sufficient for determining the fluxes, $f_\lambda$, in a set of well-defined bandpasses where $\lambda=1$ we will take to be the reference band.  If we model the measured fluxes (the data), as due to the true fluxes plus noise terms $f^d_\lambda = f_\lambda + n_\lambda$ and we assume that the $n_\lambda$ are Gaussian random variables with mean zero and  $\langle n_\lambda n_{\lambda'} \rangle = \sigma^2_\lambda \delta_{\lambda \lambda'}$ then
\begin{align}
\lefteqn{P(f^d_\lambda | t,z,m_1,\sigma^2_\lambda) = } \nonumber \\
&& \frac{1}{(2\pi)^{N/2} \Pi_\lambda \sigma_\lambda} \exp\left[-\sum_\lambda \frac{(f^d_\lambda - f_\lambda(t,z,m_1))^2}{2\sigma_\lambda^2}\right] \label{eq:lkhd}
\end{align}

We can turn this around, using Bayes' theorem, to get an expression for the probability distribution of the unknowns $t$, $z$ and $m_1$ in terms of the knowns, the $f^d_\lambda$, and whatever assumptions we care to make, which we shall call $I$ (for Information).   According to Bayes' theorem, which follows from axiomatic properties of probability distributions,
\be \label{eq:bayes}
P(t,z,m_1|f^d_\lambda,I) = P(f^d_\lambda |t,z,m_1,\sigma^2_\lambda)\frac{P(t,z,m_1|I)}{P(f^d_\lambda|I)}.
\ee
The probability distribution on the left-hand side is called the ``posterior'' distribution (because it is the probability distribution {\em after} we have the data), and the $P(t,z,m_1|I)$ is called the ``prior'' distribution (because it is the probability distribution for the parameters {\em prior} to our collection of the data).  The prior is where all our modeling assumptions are encapsulated.  The term in the denominator we can ignore since it has no dependence on the model parameters. While $m_1$ is formally a measured quantity derived from one of the band fluxes, it is treated as an unknown here because we want to map variations in $P(z)$ by perturbing it within its Gaussian noise distribution.

If we are interested in only the redshift, we can obtain its (marginal) distribution by integrating over all possible values of $t$ and $m_1$ so that
\be \label{eq:int_tm}
P(z) \equiv P(z|f^d_\lambda,I) = \int dt \int dm_1 P(t,z,m_1|f^d_I,I)
\ee
where we have written the distribution simply as $P(z)$, suppressing the dependence on the data and $I$ (modeling assumptions) for notational convenience.\footnote{Recall that $t$ may be representing an array of variables and so the integration over $t$ may be a multidimensional one.}  Although we have ignored the $P(f^d_\lambda|I)$ factor, we can now make up for that by choosing a normalization factor such that $\int_0^\infty P(z) dz = 1$.  A number of authors have used Bayes' theorem in this way to obtain the marginalized probability distribution of the redshift \citep[see for instance ][]{Benitez:1998br, Edmondson:2006fy, chapin2004}.

\subsection{Estimating $dN/dz$}

Often the redshift distribution is reduced to a single value of redshift, $z_p$, perhaps the highest peak in $P(z)$, and then this value of $z$ is referred to as a Bayesian photometric redshift estimate. Here we use $z_p$ to sort galaxies into photometric redshift bins (collections) but use all the information in the $P(z)$ for each galaxy to estimate $dN/dz$ for the collection.

\begin{figure}[t]
\includegraphics[width=0.5\textwidth]{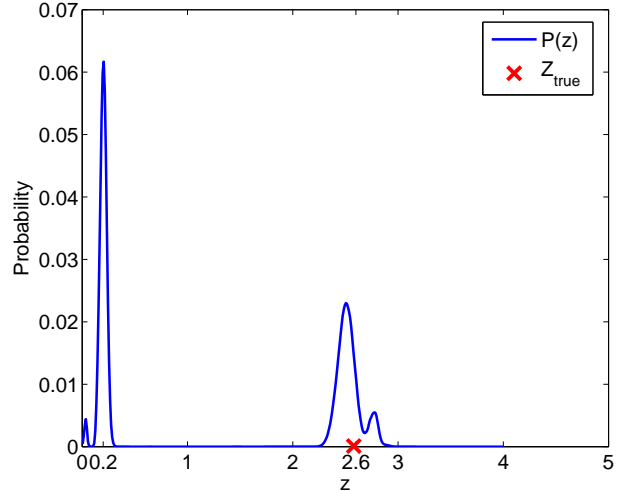}
\caption{The blue line is an example of a photometrically calculated probability distribution for a single galaxy. The true redshift (the red `x') is at $z=2.6$, whereas the $P(z)$ peaks at $.2$. By assigning non-negligible probability at the true redshift, it can be seen how the $P(z)$ can be more accurate than a single-point estimate. }\label{poz}
\end{figure}

As various authors have noted \citep{Bordoloi:2010, Cunha:2009, mandelbaum:08, Wittman:2009sv}, the use of the full $P(z)$ can improve on the systematic errors in photometric redshift estimation.  Errors in photometric redshifts are often due to degeneracies in color space. (For instance, when an object's spectrum is represented by the fluxes in low resolution optical bands, the Lyman break at $z\sim 3$ may be indistinguishable from the $4000${\AA} break at low redshift.) Although this might result in a drastically miscalculated point estimate redshift (i.e. catastrophic error), the $P(z)$ for most galaxy types near this redshift is multimodal.  Because inference of cosmological parameters (for instance through weak lensing observations) relies less on the redshift of specific galaxies and more on the \emph{distribution} of galaxies in a collection, the use of the full $P(z)$ is a vast improvement over point estimates. Consider Fig. \ref{poz}: the maximum of the $P(z)$ is at low redshift, whereas the true redshift is near $z\sim 3$.  In a point estimate approach this would be a catastrophic error, but the full $P(z)$ assigns almost half of the probability to the galaxy being near its true redshift.

Enumerating the galaxies in a given collection (redshift bin) with index $g$, we define our estimate for $dN/dz$ in that collection as
\be \label{eq:est}
\frac{dN}{dz}(z) =\sum_g P_g(z).
\ee
The estimate is motivated by the fact that the probability distribution of the redshift of a galaxy drawn at random from the collection is given by $\frac{1}{N} \sum_g P_g(z)$ where $N$ is the number of galaxies in the collection.
The Appendix\,\ref{sec:bias} proves that this intuitive estimator is not
biased as long as the prior and the likelihood in Equation (\ref{eq:bayes})
are known perfectly.

\begin{figure}[t]
\includegraphics[width=0.5\textwidth]{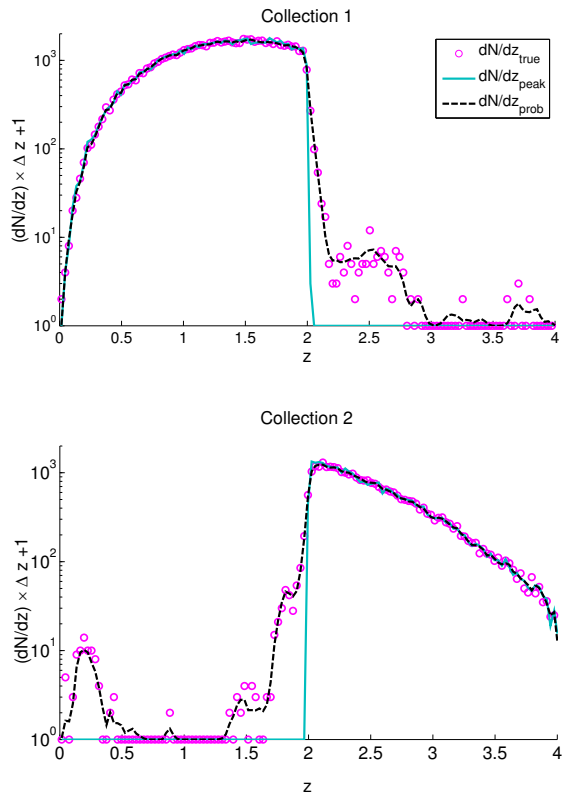}
\caption{$dN/dz$ for two redshift ``collections". Galaxies are sorted into collections on the basis of peak redshift value (or point estimate for the photo-$z$). The top plot shows collection 1, bottom plot shows collection 2. The blue line shows the $dN/dz$ estimated by sorting galaxies into collections based on their most probable redshift and summing the single point \phz estimates (the peak values of the $P(z)$).  The circles are the value of the corresponding $dN/dz$ for the true redshifts. The dashed black line shows the $dN/dz$ estimated by collecting galaxies using the peak of the probability distribution and summing the entire $P(z)$. It can be seen that $dN/dz_{prob}$ traces much more closely the true distribution than the distribution derived from the point estimate.
}\label{dndzexample}
\end{figure}

The advantage to calculating the $dN/dz$ in this way can be seen in Fig. \ref{dndzexample}, which shows the $dN/dz$ calculated for a mock galaxy catalog of $100,000$ galaxies consisting of the 6 CWWSB templates which follow the type distributions as a function of magnitude given by \citet{Benitez:1998br}(but with a fixed dN/dz peaking at z=2.0 for simplicity). The full $P(z)$ is calculated for each galaxy, and the galaxies are sorted into redshift bins, or ``collections", based on the peak value of the $P(z)$ (in other words, the photo-z point estimate). We compare the actual distribution, $dN/dz_{\rm true}$, to the $dN/dz$ calculated from summing and histogramming the single point \phz estimates, $dN/dz_{\rm peak}$. Because the calculated distribution, $dN/dz_{\rm peak}$, is constructed from the same photo-z point estimates that were used to assign each object to a collection (bin), the estimated redshift value necessarily lies within the redshift range for which the collection (bin) is defined; thus the calculated $dN/dz_{\rm peak}$ will always be zero outside that redshift range (or photometric ``redshift bin").  This is a major failing of the point estimator method since the true redshift distribution of galaxies, $dN/dz_{true}$, may have tails that extend to redshifts above and below the redshift values for which the collection is defined. By using the full $P(z)$, however, a galaxy placed in a collection can have a non-zero probability extending to values outside of redshift range of the collection.  Thus the distribution calculated using the $P(z)$ of each galaxy, $dN/dz_{\rm prob}$, can more accurately reflect the true redshift distribution.

Despite the use of the full $P(z)$ there are many possible sources of error in the \phz estimation that could still cause us to miscalculate the distribution of galaxies, $dN/dz$.  These include use of a small number of galaxies, mischaracterization of the statistical properties of the flux noise, and errors in the modeling assumptions summarized by the prior, $P(t,z,m_1|I)$.  The latter includes prior assumptions about galaxy distributions as a function of type, redshift, magnitude, surface brightness, etc., and also includes the choice of SED templates.  Errors from these modeling assumptions are the central concern of this paper.

\subsection{Parameterizing Modeling Uncertainty}
\label{subsec:PCA}

We capture the effect of modeling uncertainty by explicitly varying the modeling assumptions and seeing how the $dN/dz$'s vary in response.  We use PCA to define functions of $z$ (modes) that describe the departures from the mean $dN/dz$.   These modes, together with their amplitudes, define our parameterization of $dN/dz$.

We consider two different kinds of variation.  In one procedure we explore effects due to discrete variations in the SED templates used to generate the \phz  estimations.  The other procedure is for the case where parameters governing the prior are not perfectly known, but are constrained by auxiliary observational data.  In this latter case one can sample from the posterior distribution of these ``prior'' parameters, and generate the $dN/dz$ from this sample.

No matter how one generates the samples of $dN/dz$, the PCA is performed as follows. Given a data matrix comprised of multiple entries for a set of observables, PCA analysis finds the eigenvalues and eigenvectors of the covariance matrix for these data.  In this framework we treat the values of $dN/dz$ at discrete points in redshift space as our set of observables, and each modeling variation results in small changes in this set.  We start by forming the data matrix $\mathbf{\eta}$ for each collection
\be
\eta_{i \alpha} = \left(\frac{dN}{dz}\right)_{\alpha}(z_i).
\ee
Here $i$ indexes the points in z space for which $dN/dz$ is defined and $\alpha$ runs over the variations.  Subtracting the mean of each row from each element in the row of $\mathbf{\eta}$, we have
\be
D_{i \alpha} = \eta_{i \alpha} - \overline{\eta}_i
\ee
where
\be
\overline{\eta}_i = \frac{1}{m}\sum_{\alpha}\eta_{i \alpha}
\ee
and $m$ is the number of variations. The covariance matrix is
\be
\mathbf{C} = \frac{1}{m-1} \mathbf{DD^{\intercal}}.
\ee
Computing the eigenvalues ($\{\lambda\}$ and eigenvectors $\{U\}$, the first principal component $U_1$ corresponds to the largest eigenvalue, the second principal component $U_2$ corresponds to the next largest eigenvalue, and so on.  Each data vector (indexed by $\alpha$) can be written as a linear combination of the principal components, i.e.
\be
 \eta_{i \alpha} = \overline{\eta}_{i}+ \sum_{\beta=1}^N B_{\alpha \beta} U_{i \beta}
 \label{eq:reconstruct}
\ee
where the weights, $B_{\alpha \beta}$, are given by the linear transformation
\be
\mathbf{B}=\mathbf{U^{\intercal}\eta}
\label{eq:weights}
\ee
There are $N$ principal components of length $N$, where again, $N$ is the number of discretized values of $z$.  If the eigenvalues decrease sufficiently quickly from biggest to smallest, then most of the variation in $\eta_{i\alpha}$ is described by the first few modes.  By keeping only the first $k$ modes, we are able to reduce the dimensionality of the parameter space. We can reproduce the data via
\be
 \eta_{i \alpha}' = \overline{\eta}_{i}+ \sum_{\beta=1}^k B_{\alpha \beta} U_{i \beta}
\ee
where $\eta_{i \alpha}' \simeq \eta_{i\alpha}$ for a small value of k if the eigenvalue spectrum is
sufficiently steep.

The above description was for a single collection (``photo-z bin'') defined for a given redshift range.  It can be repeated for each of the $\gamma = 1, ... n_{z}$ photo-z collections; we will distinguish quantities for collections defined for different redshift ranges with a superscript $\gamma$.

We can now finally define our parameterization of the redshift distribution as the amplitudes of the modes $U^\gamma_{i\beta}$ defined by the above process.  We will call these amplitudes $b^\gamma_\beta$ and they influence the redshift distributions $\eta^\gamma_i \equiv dN/dz^\gamma_i$ via
\be
\eta_i^\gamma = \bar \eta_i + \sum_{\beta = 1}^k b^\gamma_\beta U_{i\beta}
\ee

Each modeling variation, $\alpha$, leads to a particular value of $b^\gamma_\beta = B^\gamma_{\alpha \beta}$ and we can use these `realizations' to model the prior distribution of the $b^\gamma_\beta$, in this lower, $k$-dimensional space.  The $b^\gamma_\beta$ are not fixed at the discrete values, but are allowed to vary continuously to describe the full space of model uncertainty. This distribution will have zero mean due to the mean removal that occurs prior to the principal
component decomposition.  We can estimate a covariance matrix via
\be
{\cal C}_{\gamma \beta, \gamma'\beta'} \equiv \langle b^\gamma_\beta b^{\gamma'}_{\beta'} \rangle \simeq \frac{1}{m-1} \sum_{\alpha = 1}^m B_{\alpha \beta}^\gamma B_{\alpha \beta'}^{\gamma'}.
\ee

\subsection{Relation to the Estimation of Cosmological Parameters}

For the analysis of cosmic shear data, we assume it is possible to take a set of model shear power spectra, $C_l^{\gamma \nu}$ and calculate its likelihood, given the shear data.  We further assume that one can take a specification of cosmological parameters, and a redshift distribution for each photo-z collection, and use these to calculate the model $C_l^{\gamma \nu}$.  From the power spectra likelihood, any prior constraints on the cosmological parameters (for example from cosmic microwave background observations), and a prior constraint on the redshift distribution parameters, one can then form a joint posterior distribution for the cosmological plus redshift-distribution parameters:
\bea
\ln P_{\rm posterior}(\theta_{\rm cos}) &= &\ln {\cal L}\left({\rm shear \ data}|C_l^{\gamma \nu}(\theta_{\rm cos},b^\gamma_\beta)\right) \nonumber \\
& + & \ln P_{\rm prior}(\theta_{\rm cos}) + \ln  P_{\rm prior}(b^\gamma_\beta)
\eea
where
\be
\ln  P_{\rm prior}(b^\gamma_\beta) = b_{\gamma \beta} {\cal C}^{-1}_{\gamma \beta,\gamma'\beta'} b_{\gamma' \beta'}/2
\ee
With the ability to calculate this joint posterior,  one would then be able to explore the constraints on this parameter space given by the data via Markov Chain Monte Carlo, for example.

\section{Example 1:  Template Uncertainty}
\label{sec:templatenoise}

One of the potentially largest sources of modeling errors in photometric redshift estimation may come from the selection of the template set in the SED fitting method, the technique currently used for deep and faint redshift surveys. At present, many \phz codes use a small number of template spectra from nearby galaxies e.g. the popular CWW+SB set \citep{CWW:80,Kin:96} used in part of our analysis. Though the SEDs are sometimes `tweaked' to better fit the data, \citep[e.g. ][]{Budavari}, expanding beyond such simple template sets almost always involves the use of population synthesis models \citep{EAZY,COSMOS}, due to the lack of sizeable numbers of high signal-to-noise spectra at high redshift. The difficulty of spectroscopic followup of faint, high redshift galaxies means these models may be incomplete and unrepresentative of true galaxy evolution.  Such biases can have a significant effect on the \phz estimations \citep[see for instance ][]{MacDonald:2010eu}.  Because the SED shape for some population of galaxies may evolve, or a population not represented at low redshifts/high luminosities is present in the deep photometric data, the choice of template set is an important source of modeling uncertainty.

There are also convergence issues. Since the \phz is a nonlinear inverse problem, a small amount of photometry noise will drive a noisy assignment of templates for each galaxy.  The amount of this induced template noise is a nonlinear function of the photometry noise.  This is a general feature of threshold systems: the signal-to-noise ratio peaks at a small non-zero value of input noise \citep{Wiesenfeld:1995sp}.  At zero noise the \phz error is large and due entirely to errors in the modeling assumptions, whereas at high noise levels the \phz error is large because no template assignment converges. In this paper we focus on \phz errors due to errors in models of SED templates and magnitude and type priors at constant photometry noise given by the LSST data model.

To begin to explore the sorts of uncertainties present in an actual survey with noise and where galaxy types may be left out or incorrectly modeled, we start by generating a mock catalog based on 20 SED templates. We then calculate the photometric redshifts while systematically leaving out one each of the SED's at a time. When one of the templates is removed, the galaxies in the catalog matching this SED will be fit to another, incorrect, SED, creating errors in the photometric redshifts and variations in the resulting $dN/dz$. The result is a set of $21$ distinct photometric redshift estimates for the galaxies in the catalog, which we can use to develop a parameterization and model of uncertainty for the redshift distribution. (We note that this is just one of the many ways in which template noise can be examined.)

\subsection{Mock Data}

To generate our set of templates, we begin with a catalog of galaxies from GOODSN \citep{Giavalisco:2003ig} with measured spectroscopic redshifts and seventeen band photometry, covering U-band to IRAC mid-infrared wavelengths, available as part of the Photo-\emph{z} Accuracy and Testing (PHAT) program \footnote{Further information on PHAT,
including the dataset used here is available at: http://www.astro.caltech.edu/twiki\_phat/bin/view
/Main/WebHome} \citep{Hild10}.  Although there are hundreds of galaxies in the sample, for tractability we develop a more manageable set of 20 SED templates that are representative of the larger dataset.

\begin{figure}[t]
\includegraphics[width=0.45\textwidth]{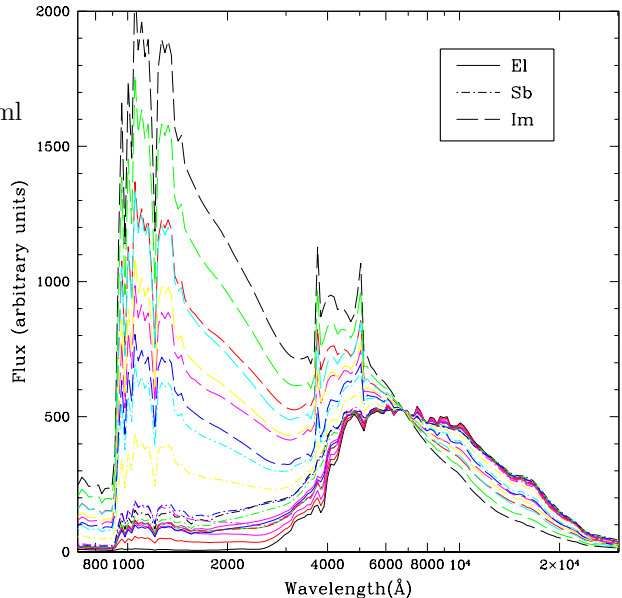}
\caption{The 20 LRT SED templates derived from GOODSN and linear combinations of the Assef basis templates. The galaxies are further broken into three type classifications which follow the redshift distributions as parameterized in Table \ref{tab-priorpars}.} \label{fig:LRTtemps}
\end{figure}

We derive our template set using the methodology of \citet{Assef:2007qy, Assef:2010jm} in which the 17 band data for each galaxy is fit with a non-negative linear combination of four basis templates (Elliptical, Spiral, Irregular, and active galactic nuclei (AGN)) shifted to the measured redshift.  Modifying the code available at http://www.astronomy.ohio-state.edu/~rjassef/lrt/index.html to allow for the extraction of the coefficient associated with each component of each component and eliminating galaxies with a strong AGN component, we arrive at a sample of 375 galaxies, each described by a linear combination of E, Sbc, and Im templates. Examination of this reduced three dimensional space reveals that the galaxies' coefficients are not significantly clustered.  In order to choose a more manageable, but still representative, set of templates that spans the dataset, we run a simple k-means algorithm \citep{macqueen67} to choose 20 sets of coefficients that will comprise our mock galaxy template set. The resulting templates are shown in Fig. \ref{fig:LRTtemps}. To facilitate the following step, we categorize the templates in terms of their dominant morphological type, E, or $T_1$, Sbc or $T_2$ and Im, $T_3$, with the reddest templates $1-6$ corresponding to $T_1$, templates $7-13$ to $T_2$ and the bluest templates $14-20$ corresponding to $T_3$.  

We generate magnitudes for the LSST $ugrizy$ filter set, with depth and error properties expected for the ``gold'' sample described in section 3 of the LSST Science Book \citep{lsst:2009}, which consists of expected full ten year depth and cut to $i<25.0$ magnitude.  The $5\sigma$ limiting magnitudes in the $ugrizy$ bands are: 26.1, 27.4, 27.5, 26.8, 26.0, 24.8.  The ``gold'' sample galaxies will have S/N $\gtrsim 20$ in multiple filters, which is highly recommended for meaningful photo-z measurements. We begin by generating an \emph{r}-band apparent magnitude that follows $p(m)\,\propto\,10^{0.37m}$ to approximate a typical number count distribution.  We then choose one of the 20 template types for each galaxy by Monte Carlo, such that the morphological types, $T_1$, $T_2$ and $T_3$ are distributed according to
\be
\label{eqptm}
p(T|m_0) = f_t \exp\{-k_t(m_0-20)\}.
\ee
We then assign via Monte Carlo a redshift to each galaxy, but such that
\be
p(z|T,m_0)\propto z^{\alpha_t}\exp\{-[\frac{z}{z_{m_t}(m_0)}]^{\alpha_t}\}
\label{eqpzdndz}
\ee
is satisfied, where $z_{m_t}(m_0) = z_{0_t} + k_{m_t}(m_0 - 20)$, and $T$ refers to E, Sbc, or Im. The free parameters, $f_t$, $k_t$, $z_{0_t}$ and $\alpha_t$, are fit to a training set of spectroscopically measured redshifts. We take these to be the values given in \citet{Benitez:1998br}, (listed in Table \ref{tab-priorpars}). Given the redshift, type, and \emph{r}-band magnitude, the expected magnitudes in the remaining $ugizy$ bands are assigned appropriately.

\begin{table}[ht]
\centering
\caption{Fiducial parameter values for magnitude priors}
\begin{tabular}{|l l l l l l|}
\hline
Spectral type & $f_t$  &$\alpha_t$ & $z_{0t}$ & $k_{mt}$ & $k_t$ \\
	\hline \hline
E/SO & $0.35 $&$2.46$ & $0.431$ & $0.0913$ & $0.450$ \\
Sbc,Scd & $0.50 $&$1.81$ & $0.390$ & $0.0636$ & $0.147$ \\
Irr & $0.15 $&$0.91$ & $0.0626$ & $0.123$ &  \\
	\hline
\end{tabular}
\label{tab-priorpars}
\end{table}

\begin{figure}[t]
\includegraphics[width=0.5\textwidth]{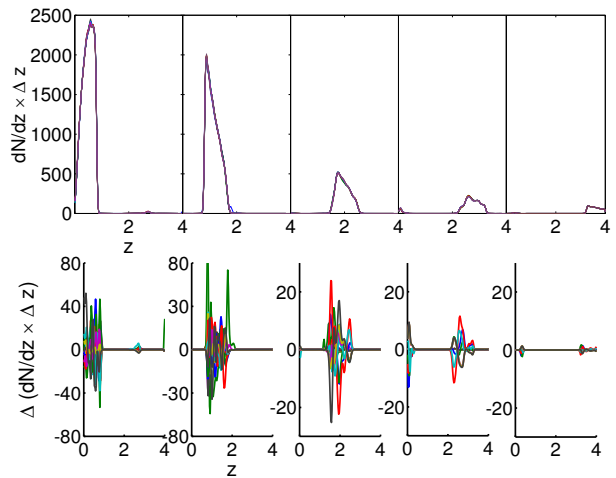}%
\caption{LRT template sets: top panels show the $dN/dz$'s from summing the $P(z)$ of galaxies in each collection defined for a given redshift range.  The bottom panels show the residuals for each realization -- that is the difference between the fiducial (all 20) template set and the reduced (19 only of the 20) template sets, $\Delta dN/dz_{i} = dN/dz_{fid} - dN/dz_{i}$, for each template set realization.}\label{fig:LRTdndzresids}
\end{figure}

\subsection{Results}

With the fiducial catalog generated, we can begin to explore how template noise affects the \phz distribution. We run the \phz estimation code BPZ \citep{Benitez:1998br} to calculate the $P(z)$ for each galaxy in the catalog.  We perform 21 iterations -- once with the fiducial model, e.g. all 20 SED's included for the estimation, and then 20 variations, each realized by leaving out one of the 20 SED templates.

Each incomplete set of SED's produces slightly different errors in the \phz estimations, which result in variations in the $dN/dz$ for each collection. The $dN/dz$ is stored as a vector of $134$ ``microbins" with a width of $.03$ from $z=0$ to $z=4$. Hence the sum,
\be
N_{\gamma} = \sum_{j=1}^{134} dN/dz_{j\gamma},
\ee
returns the total number of galaxies per collection $\gamma$, and summing over the galaxies in each collection returns the total number of galaxies in the catalog, e.g. $100,000 = \sum_\gamma N_\gamma$.  In Fig. \ref{fig:LRTdndzresids} the $dN/dz$ are shown for each realization of the templates.  The bottom panels show the residuals, $dN/dz_{i} - dN/dz_{all}$, where $i$ refers to a realization where a single template has been left out for the \phz estimation.  It can be seen that the residuals are on the order of a few percent when compared to the $dN/dz$.

\begin{figure}[t]
\includegraphics[width=0.5\textwidth]{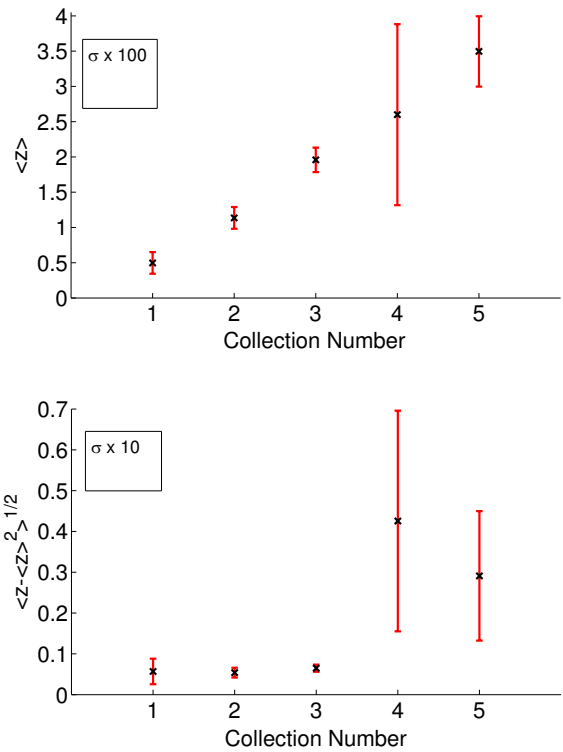}
\caption{LRT set: Top panel shows the average of the bias, $\mu_{<z>} - z_{center}$, where $z_{center}$ is the center of the redshift range for the collection. (This is essentially a plot of the average of the mean redshift for each variation, but the subtraction of the collection center allows the scale of top and bottom plots to be the same.) The standard deviation (spread) in the bias is plotted as error bars, where for visibility we multiply by a factor of 100: $\sigma \times 100$.  The bottom panel shows the average variance, $\mu_{\Delta}$, and the standard deviation in the variance is plotted as error bars.  For visibility we multiply by a factor of 10: $\sigma \times 10$. The large error bars in collections 4 and 5 show the effects of the degeneracies present in faint galaxies at low redshift.
}
\label{fig:stndv_errorbars}
\end{figure}

The distribution of galaxies can be described by the mean and variance of the $dN/dz_i$ per collection $\gamma$, and are different for each template set realization. (Note that the collection index $\gamma$ has been suppressed for clarity.)
\be \langle z \rangle_i = \int dz ~ z ~ dN/dz
\label{eq:mean}
\ee
and
\be
 \triangle_i = \sigma^2 = \int dz (z- \langle z \rangle)^2 dN/dz
 \label{eq:stdv}
\ee
where $i$ refers to the specific template choice. We calculate the mean and the standard deviation of the previous quantities, i.e.
\be
\mu_X = \frac{1}{21}\sum_{i=1}^{21}(X_i)
\ee
\be
\sigma_X = \sqrt{\frac{1}{21}\sum_{i=1}^{21}(X_i - \overline{X})^2}
\label{eq:variance}
\ee
where $X$ refers to either the mean or the variance of each variation.
These values are shown in Fig. \ref{fig:stndv_errorbars}.  (To keep the axes on the same scale, we actually plot the average offset for the bin, $\mu_{<z>} - z_{center}$.)

The spread in the means is typically on the order of $10^{-3}$, although for collection $4$ it is larger.  This reflects the difficulties in calculating photometric redshifts for the redshift range for collection $4$ is defined, $2.4<z<3.2$  Specifically, the Lyman break at $z \sim 3$ is easily confused with the the $4000${\AA} at low redshift. (This is the cause of the low redshift ``bump'' in the $dN/dz$ in collection $4$ seen in Fig. \ref{fig:LRTdndzresids}.) This effect is even more apparent in the standard deviation of the variance for collection $4$.  The variance is sensitive to probability moving into and out of outlying redshift ``islands". To a lesser extent, this effect is also visible in the spread of the variance for collection $1$.  The effect is smaller, however, due to the large numbers of bright galaxies that also populate collection $1$ and are never confused with distant galaxies. (Note the asymmetry of this degeneracy, which makes this effect more pronounced in the higher redshift bins: bright objects are unmistakably at low redshift, whereas faint objects may either be low redshift intrinsically faint galaxies or high redshift intrinsically bright galaxies.)

\begin{figure}[t]
\includegraphics[width=0.5\textwidth]{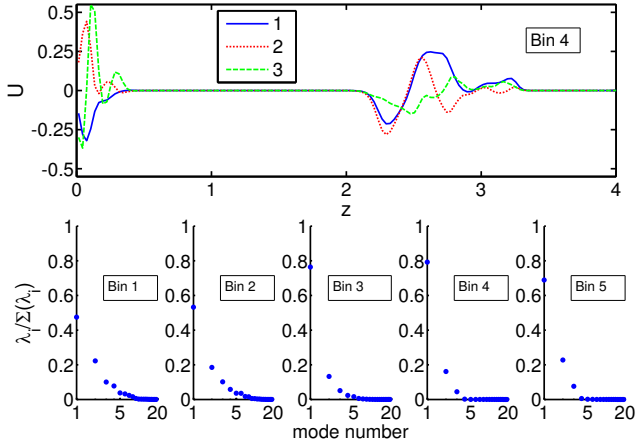}
\caption{LRT set: Top panel shows the principal components for collection $4$ (illustrative of principal components for other collections, here simply labeled ``bins''). The bottom panels show the eigenvalue spectrum (i.e. amount of variance captured by each principal component) (x-axis is logarithmic). In all cases, 80\% or more of the variance is capture by the first three modes.} \label{fig:pcs_eigens_LRT}
\end{figure}

We calculate the principal components of the $dN/dz$ due to the variations of the template sets.  The fraction of the variance captured by a given mode $i$ is $\lambda_\alpha\sum_{i}\lambda_i$, where $\lambda_i$ is an eigenvalue of the covariance matrix corresponding to the $i$th mode.  It is difficult to assign a physical interpretation to the modes themselves, which show a complex and data-dependent structure. The modes will be a function of the methodology used to vary input model, e.g. we could have chosen to leave out 2 templates at a time and the modes returned would have shown a different structure; however, they will always represent the ``directions" in the data, rank-ordered by the amount of variance captured. As an example, the first three modes of collection $4$ are plotted in Fig. \ref{fig:pcs_eigens_LRT}. Some features of the underlying data can be seen in the modes. The high amplitude of the modes at low redshift, for instance, reflects the fact that the template set variations move probability of certain galaxies into and out of this low redshift outlying region.

\begin{figure}[t]
\includegraphics[width=0.5\textwidth]{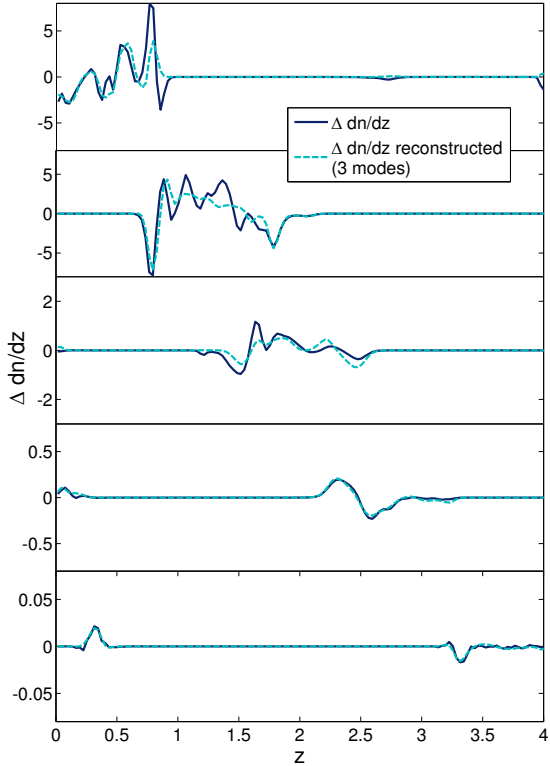}
\caption{LRT set: Depicts $d_{all}$: the mean-subtracted $dN/dz$ for the case of all 20 SED templates. The blue (solid) line is the the actual residual. The cyan (dashed) line is the residual reconstructed from the first three principal components.   Although some of the details (wiggles) are missed, the general shape is reproduced.} \label{fig:LRT_resid_recon}
\end{figure}

\begin{figure}[t]
\includegraphics[width=0.5\textwidth]{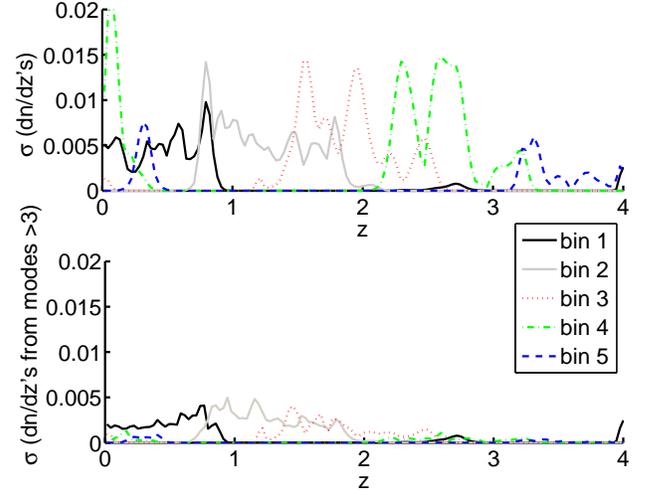}
\caption{LRT set: The top panel shows the standard deviation of the variations $d_{\alpha}$ (the mean-subracted $dN/dz_{\alpha}$) at each point in redshift space. The bottom panel is the standard deviations of the $d_\alpha$ reproduced from linear combinations of modes 4 and higher.  It can be seen that most of the variance can be accounted for with modes $1-3$ -- little is described by the higher number modes.}
\label{fig:stndv_ofz}
\end{figure}

As noted above, the principal components are useful for dimensionality reduction.  For all collections, the first three principal components cumulatively account for 80\% or more of the variance.  The eigenvalue spectrum is plotted in Fig. \ref{fig:pcs_eigens_LRT}, where it can be seen that higher number modes quickly approach zero.

Recall from Subsec. \ref{subsec:PCA} that the first step of PCA is to subtract the mean from each $dN/dz_{\alpha}$, where $\alpha$ refers to a specific realization of template set. We define $\{d_\alpha\}$ to be the set of mean-subtracted data, i.e. $dN/dz_\alpha - \langle dN/dz \rangle$.
Each $d_\alpha$ can be reconstructed from a specific linear combination of principal components. Performing the reconstruction of the $dN/dz$'s of the LRT template variations with the first three modes, it is possible to recover 80\% or more of the variance per collection. In general, the precise amount of variance recovered with the first three modes will depend on the specific methodology used to vary the \phz modeling assumptions and also on the data set.  In general, the variance and the fidelity of the reconstruction can be ``tuned" by retaining a larger or smaller number of the modes, e.g. if $1\%$ accuracy is required, we can retain seven or eight modes that capture $99\%$ of the variance.  In the case of three modes retained and five tomographic bins, the \phz uncertainty is parameterized by just $15$ free parameters. This is a relatively tractable number considering analyses propagating redshift uncertainties through to the cosmological parameters have used $60$ or more free parameters \citep{Ma:2005rc,Zhan:2006gi,Bernstein:2007uu}.

A potential concern is that there is a mode (or modes) of dN/dz, discarded by the truncation,
which, though not contributing much to variance of dN/dz, causes a change to the shear
power spectrum (or other observable) with the same shape as caused by a cosmological parameter variation.
Thus, although the amplitude of the variation is small, it could lead to bias in parameter
estimation.  We think such a scenario is unlikely, but one could test for it when this
work is extended all the way to cosmological parameter estimation.

Additionally we note that knowledge of the specific redshift range of dN/dz to which a cosmological parameter is most sensitive could improve the mode selection. Our PCA modes are defined and ordered based on the amount of variance they account for in $dN/dz$. Of course the variation in the dN/dz is a function of redshift. For our analysis, we weight variations happening at different redshifts equally; however, one might improve the mode decomposition and their ordering by using some redshift-dependent weighting of the variation. We expect the optimal weighting to be a slowly-varying function of redshift, such as for example, a downweighting variations at $z=4$ relative to those at $z=1$. It may be worth exploring such an optimization in the future.

Fig. \ref{fig:LRT_resid_recon} shows $d_{all}$, the mean-subtracted $dN/dz$ for the case of all 20 SED templates.  The blue (solid) line is the actual residual and the cyan (dashed) line is the reconstruction from the first three principal components. While some of the detailed structure is missed, the overall shape is reproduced. The addition of more modes would continue to improve the reconstructed $d_{all}$.

What about the variance of the variation? In the top panel of Fig. \ref{fig:stndv_ofz} we plot the standard deviation of the set of $d_\alpha$ at each point in redshift space. This essentially captures the uncertainty in the photometric redshift calculations as a function of redshift.  However, it is not necessary to retain all the modes to capture this level of error; rather with just three modes over 80\% of this uncertainty can be represented. This is illustrated in the bottom panel of Fig. \ref{fig:stndv_ofz} where we reconstruct the $d_\alpha$ only using modes 4 and higher and recalculate the standard deviation as a function of $z$ using these reconstructed $d_\alpha$.  It can be seen that the standard deviation from the reconstruction is greatly reduced for all collections -- in other words, a large portion of the variance is captured by the first three principal components and little is accounted for by the higher order modes.

\section{Example 2:  Magnitude and type prior distribution uncertainty}
\label{sec:magpriors}

Another modeling assumption that can affect the photometric redshift estimations is the assumed distribution for redshifts as a function of type and magnitude.  In the Bayesian framework, these distributions can be applied as a prior in the SED template fitting method with the intent to improve photometric redshift estimates. In \citep{Benitez:1998br} these priors are derived semi-empirically, that is to say, a functional form is chosen and the parameters are fit with a training set of spectroscopic redshifts. The probability of a galaxy being at a certain redshift is given by
\be
p(z|C,m) \propto \sum_{T}p(z,T|m_0)p(C|z,T)
\ee
and using the product rule
\be
\label{eqpztm}
p(z,T|m_0)= p(T|m_0)p(z|T,m_0).
\ee
The distributions, $p(T|m_0)$ and $p(z|T,m_0)$ are constrained using auxiliary data. Errors may be introduced into these priors in a number of ways. For instance, a biased spectroscopic redshift sample (one that underrepresents a certain galaxy population present in the photometric sample) could cause errors in the inferred type distributions. An incorrect analytical form could also be chosen that is unable to accurately represent the prior distributions; or simple statistical errors, such as those resulting from the finite training set, could cause the prior parameters to be known only to low accuracy.  In this section, we show again that PCA can be used to describe uncertainties in redshift estimates resulting from such errors in the magnitude and type priors.

\subsection{Mock Data}

To investigate the effects of imperfect knowledge of the Bayesian prior, we generate a mock ``training'' set of $100,000$ galaxies.  The simulated data are created according to the procedure described in \S\,\ref{sec:templatenoise}: the generating \emph{r}-band magnitudes, types and redshifts via Monte Carlo, while following the distributions for $p(T|m_0)$ and $p(z|T,m_0)$, given in Equations \ref{eqptm} and \ref{eqpzdndz}, with the free parameters of the distribution given in Table \ref{tab-priorpars}.

Instead of using the LRT template set as a basis for generating the 17 band photometry, we use the ${\rm{CWW\!+\!KIN}}$ set that are included with the BPZ software, assigning elliptical/reddest as type 1 and SB2/bluest as type 6.  Rather than limiting ourselves to an overly simplified model with only six discrete types, which would result in very few type misidentifications, we create a continuous color distribution by linearly interpolating between the galaxy types (a continuous number between 0.8 and 6.2) to assign color to each galaxy. 

\subsection{Results}

We use the mock catalog as a starting point for probing the uncertainty in the prior parameters, $f_t$, $k_t$, $z_{0_t}$ and $\alpha_t$, and investigating the effects of this uncertainty on the photometric redshift estimations. We run BPZ to fit each galaxy to one of the standard 6 HDFN templates. (Because of the interpolation step (described above), some galaxies will necessarily be mistyped and these errors, on top of statistical errors associated with a finite training set, could further affect the estimation of the prior parameters.)

With this output, we use the Monte Carlo Markov Chain (MCMC) method (Metropolis Hastings algorithm) to find the best fit values and the confidence bounds on $f_t$, $k_t$, $z_{0_t}$ and $\alpha_t$. (Note that although there are many estimators that may have worked in our case, MCMC was chosen because of the high dimensionality of the parameter space: 14 free parameters.) The results are presented in Table \ref{tab-fitpars}.  It can be seen that these values are close to, but not the same as, the fiducial parameters in Table \ref{tab-priorpars}.

\begin{table}[ht]
\centering
\caption{Parameter values for fit using MCMC for magnitude and type priors}
\begin{tabular}{|l l l l l l|}
\hline
Spectral type & $f_t$  &$\alpha_t$ & $z_{0t}$ & $k_{mt}$ & $k_t$ \\
	\hline \hline
E/SO    & $0.305 $&$2.350$ & $0.416$ & $0.0935$ & $0.368$ \\
Sbc,Scd & $0.503 $&$1.640$ & $0.356$ &  $0.070$ & $0.152$ \\
Irr     & $0.192 $&$0.902$ & $0.0687$ & $0.119$ & $$ \\
	\hline
\end{tabular}
\label{tab-fitpars}
\end{table}

To probe the effects of variations in the priors, we select $100$ values of the parameters drawn from the Markov chain.  To make sure we consider both best and worst case scenarios, we also add the best fit values and parameter values at the upper and lower $2\sigma$ confidence levels (note that when \textit{all} the parameters are at the $2\sigma$ confidence limit, this reflects a much larger than $2\sigma$ fluctuation.)

We run the \phz redshift code, BPZ, on our simulated galaxy catalog for each instance of prior parameters.  As in the template uncertainty case, for each galaxy in our catalog, the product of the code is a probability distribution, $P(z)$, for $0<z<4$.

As for the LRT template variations, we sort the galaxies into collections by defining 5 redshift ranges evenly spaced from $z=0$ to $z=4$ and by using the criteria that the peak of the $P(z)$ fall within one of these ranges to assign each galaxy to a given collection or "redshift bin". As for the previous case, the $dN/dz_i$ for each collection, $i$, is calculated by summing the $P(z)$ for each galaxy in the collection.

\begin{figure}[t]
\includegraphics[width=0.5\textwidth]{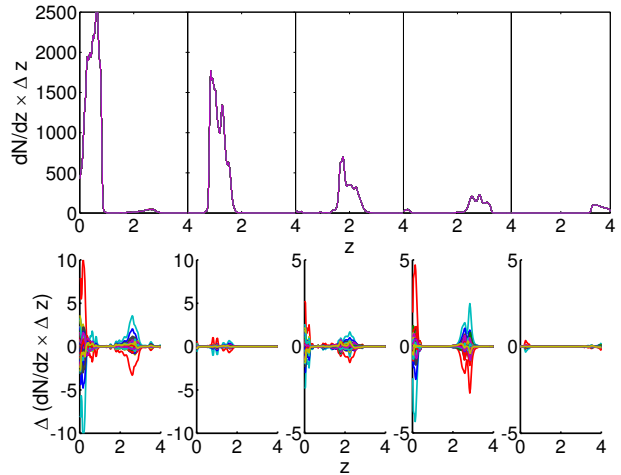}
\caption{Continuous HDFN priors: Variations in prior parameters drawn from Markov chain. The top panels show the $dN/dz$ per collection (tomographic redshift bin), the bottom panels show the difference between the $dN/dz$ from the best fit values and the $dN/dz$ from the variations, $dN/dz_{bf} - dN/dz_{i}$, for each collection. These differences are small in comparison to those of the LRT template variations.}
\label{fig:dndz_resids_cont}
\end{figure}

Variations in the priors create changes in the estimated $dN/dz$.  The $dN/dz_i$ and the residuals, $dN/dz_i - dN/dz_{\textrm{best-fit}}$ is plotted in Fig. \ref{fig:dndz_resids_cont}.  The differences between the overall $dN/dz$'s plotted here and those of the LRT model (shown in Fig. \ref{fig:LRTdndzresids}), are due to the mock data and the template set used for the \phz estimation (not the priors).  While we avoid the overly simplistic 6 template (CWW+SB) model by generating data to include linear combinations of the SEDs, we only compare to the six discrete HDFN templates to estimate the redshifts.  Thus, not every galaxy is accurately represented in the template library.  It is the non-representativeness of these templates that cause the deviations seen in Fig.~\ref{fig:dndz_resids_cont}. (Note that in the LRT case, we left one template out of 20 instead of using 6 to represent a continuous mix of types.) The variations in the $dN/dz$ due to variations in the prior are small, as can be seen by the difference between the $dN/dz$ using the best fit prior parameters and the $dN/dz$'s from each Monte Carlo sampling $dN/dz_i - dN/dz_{\textrm{best-fit}}$.  They are a factor of 5 or more smaller when compared to the variations due to the template uncertainty.  Thus, at least for the case of a well-sampled training set, small errors in the prior are not as important as having well-calibrated SED template set.

\begin{figure}[t]
\includegraphics[width=0.5\textwidth]{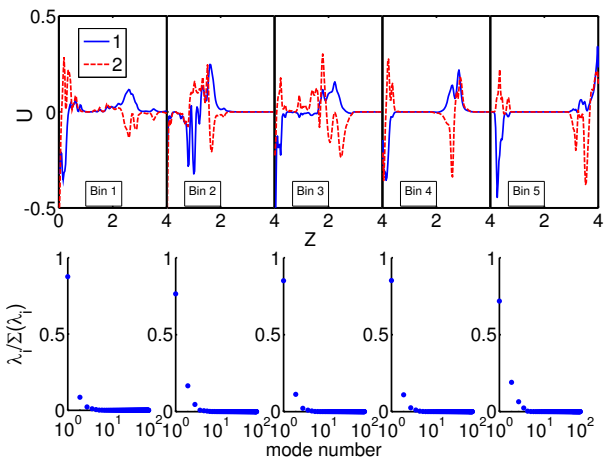}
\caption{Continuous HDFN priors: Top panels of the plot show the first two principal components for each collection. The bottom panels show the eigenvalue spectrum (i.e. amount of variance captured by each principal component). In all cases, more than 80\% of the variance is captured by the first 2 principal components.}
\label{fig:pc_eigs_cont}
\end{figure}

Once again, we perform PCA on the set of $dN/dz$ to derive a lower dimensional parameterization of the \phz uncertainties.  We find that for variations of the priors, only two modes are necessary to account for over 80\% of the variance. The top panels of Fig. \ref{fig:pc_eigs_cont} show the first two principal components for each collection.  The bottom panels show the eigenvalue spectrum for each collection.

One caveat to this approach is that we have not included the differences in the $dN/dz$'s resulting from the difference between the \emph{fiducial} prior parameters and the \emph{best fit} parameters.  However these differences in the $dN/dz$ remain small and are of the same order as the variations we do consider.  We note that in a case where the type distributions of the training set are more drastically biased, the prior parameter distribution may have a greater impact.  What we have presented here is a method for exploring such uncertainties and parameterizing their effects on the redshift distributions in such a way that they could be propagated through to error bounds on cosmological parameters.

\section{Connection to Other Work}
\label{sec:otherwork}

Many authors have considered the propagation of \phz errors through to the cosmological parameters, see for instance \citet{Huterer:2004tr, Ma:2005rc,  DETF2006, Zhan:2006gi, Abdalla:2007uc, Bridle:2007ft, Bernstein:2007uu}.  A commonly quoted result is from \citet{Ma:2005rc}, where it was shown that for a \phz distribution modeled as a Gaussian with redshift dependent mean and standard deviation, that for the a two parameter dark energy model, (the dark energy density $\Omega_{\rm DE}$, its equation of state today $w_{0} = p_{\rm DE}/\rho_{\rm DE}|_{z=0}$), the $ 1\sigma$ errors on $z_{\rm bias}(z)$ and $\sigma_z(z)$ must be less than $.01$ per ``bin'' in order to avoid a degradation in the dark energy parameters of more than $50 \%$.  Note that this result \emph{does not} apply to the mean and bias for the tomographic redshift bins (collection in the terminology of this paper), rather the analysis refers to the mean and bias on ``micro bins" of width $\delta z = .1$. Moreover, it does not include the powerful self-calibration enabled by a joint analysis of WL and BAO, due to the shared large scale stucture.

To put in perspective the variations we produce via changes in template sets and magnitude priors, we compare the size of the resulting variations in the $dN/dz_\gamma$ per collection $\gamma$ to those of the model analyzed in \citet{Ma:2005rc}. The $dN/dz_\gamma$ per collection is given by
\be \frac{dN}{dz}_\gamma(z) = \int_{z_{\rm{ph}}^{(\gamma)}}^{z_{\rm{ph}}^{(\gamma+1)}}
dz_{\rm{ph}} \, \frac{dN}{dz}_{tot}(z)\, p(z_{\rm{ph}} | z)\,.
  \label{eq:ni}
\ee
where
\be
\frac{dN}{dz}_{tot} \varpropto {z}^{\alpha} \exp\left [-(z/z_0)^\beta\right ] \,
\label{eq:nz}
\ee

The \phz distribution is modeled as
\be
p(z_{\rm ph} | z) =
  {1 \over {\sqrt{2\pi} \sigma_z}}
  \exp\left [-{{(z-z_{\rm{ph}} -z_{\rm bias})^2} \over {2 {\sigma_z}^2}}\right ]
  \label{eq:f}
\ee
and is defined for ``micro" bins of width $\delta z = .1$, each with two \phz parameters the bias, $z_{\rm bias}(z)$, and the scatter, $\sigma_z(z)$.  The $dN/dz_\gamma$ for each ``macro" bin, or collection, is obtained by interpolating the between each $p(z_{\rm ph} | z)_i$, to return a continuous function defined for the entire redshift range, and then integrating Eq. \ref{eq:ni}.

To calculate the fluctuations in each collection resulting from $1\sigma$ errors of $.01$ on $z_{\rm bias}(z)$ and $\sigma_z(z)$ for each ``micro" bin, we start with the \phz model described above.  We assume a fiducial case with $40$ ``micro" bins between $z=0$ and $z=4$ with $\delta z (z) = .1$, $z_{\rm bias}(z)= 0$ and $\sigma_z(z) = .05(1+z)$.  We then randomly draw 21 samples (consistent with the fact that we generated 21 samples from variations of template set (we could have used $103$ to correspond to variation of the prior parameters, but given that the distribution is Gaussian, the extra variations add little information) from a 80-D Gaussian distribution with $\sigma_{bias} = \sigma_{scatter} = .01$. We linearly interpolate to create a continuous function in $z$, $P(z_{\rm ph} | z)$, and we perform the integral in equation \ref{eq:ni} each time to obtain a set of $21$ $dN/dz_{\alpha}$ for each collection.

\begin{figure}[t]
\includegraphics[width=0.5\textwidth]{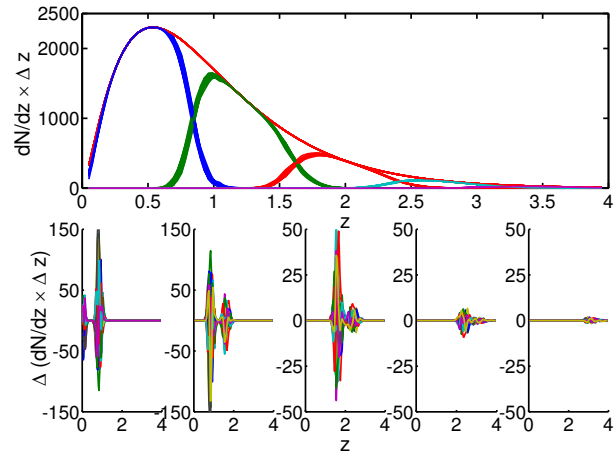}
\caption{Gaussian model \citep[from ][]{Ma:2005rc} with $\delta \sigma = \delta \mu = .01$ The top panel shows the $dN/dz$ per collection (tomographic redshift bin), the bottom panels depict the residuals, $dN/dz_{fid} - dN/dz_{i}$, in each collection. Despite stringent constraints on the $\delta\sigma$ and $\delta\mu$ for each ``microbin", the residuals in the $dN/dz$ resulting from variations at this level are larger than those of both model-based variations we consider as shown in Fig.
\ref{fig:LRTdndzresids} and Fig. \ref{fig:dndz_resids_cont}.  Catastrophic failures with significant probability well outside their expected collection are also absent from this parameterization.  }
\label{fig:dndz_resids_gauss01}
\end{figure}

\begin{figure}[t]
\includegraphics[width=0.5\textwidth]{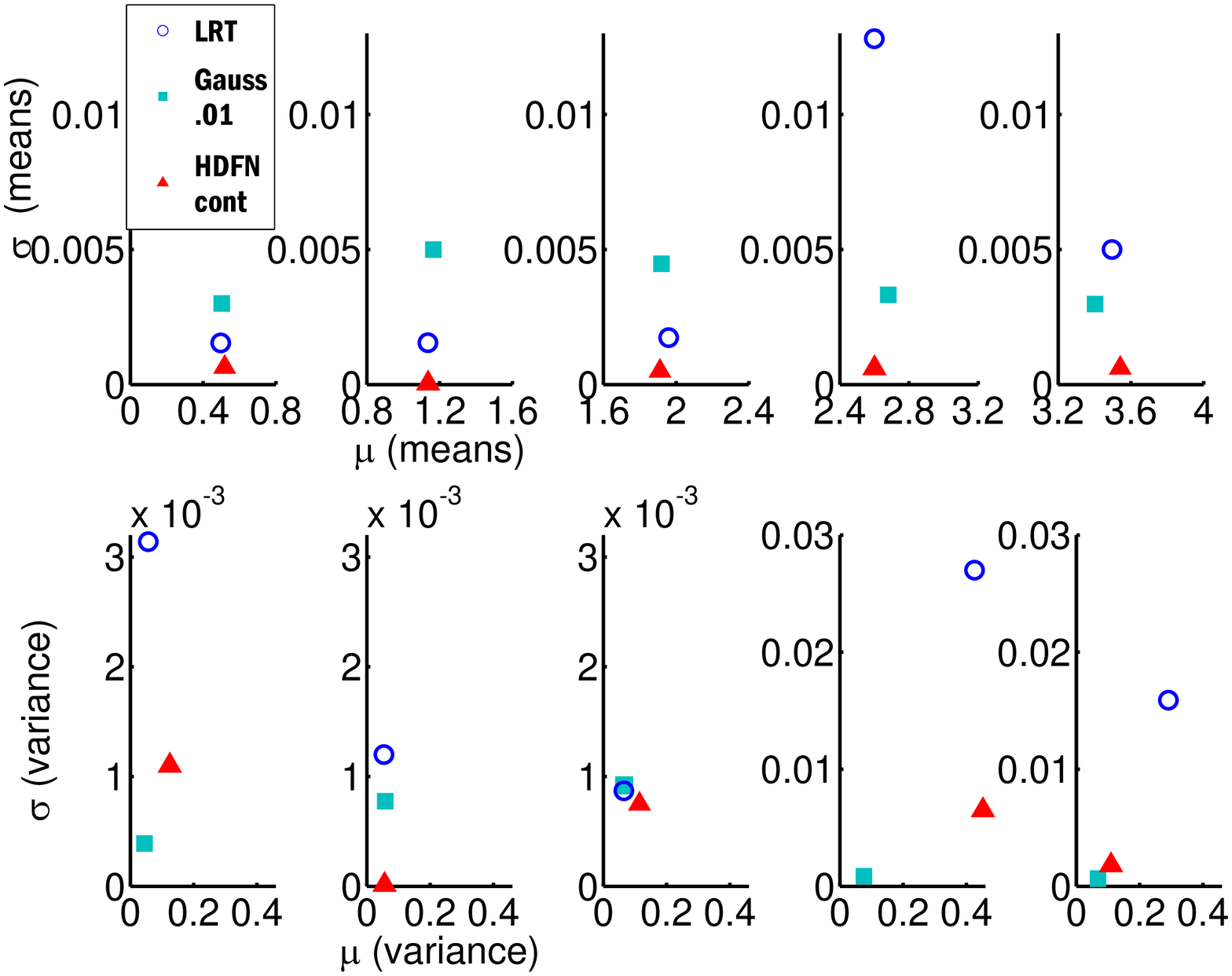}
\caption{Comparison of the standard deviation of the mean and the variance of the $dN/dz_i$, the redshift distribution per collection, for the three models considered (LRT, priors (Continuous HDFN) and Gaussian with priors on mean and spread of .01).  It can be seen that the variance in the means for the Gaussian model is often larger than that of the model-based variations. The larger variance in the LRT variations and the prior variations for collections 4 an 5 is indicative of catastrophic outliers which the Gaussian parameterization cannot represent.}
\label{fig:stdvs_compare}
\end{figure}

We define, as in \S\,\ref{sec:templatenoise}, $d_\alpha = dN/dz_\alpha - \langle dN/dz \rangle$, i.e. the mean-subtracted data.  The $dN/dz_\alpha$ and the $d_\alpha$ are shown in Fig. \ref{fig:dndz_resids_gauss01}.  It can be seen that $d_\alpha$ are of the same order (although somewhat larger) than those of the template variations (Fig. \ref{fig:LRTdndzresids}).  However, there are no catastrophic ``islands", i.e. non-zero values for the $dN/dz$ far outside of the redshift range for which the collection is defined.  This is also reflected in Fig. \ref{fig:stdvs_compare} which shows the standard deviation in the mean redshift (Eq. \ref{eq:mean}) and the variance (Eq. \ref{eq:stdv}) for each collection.  What is perhaps most interesting to note is the difference between the Gaussian case and the model-based variations in collections 4 and 5, and to a lesser extent collection 1, which are subject to the redshift degeneracies associated with faint galaxies, as discussed in \S\,\ref{sec:templatenoise}. In particular, these differences are apparent in the standard deviation in the variances (bottom panels of Fig. \ref{fig:stdvs_compare}), where the values are significantly higher for the LRT template set variations and even, though to a lesser extent, for the variations the HDFN priors.

Given the differences between the Gaussian case and our model-based variations, we refrain from drawing any conclusions about what this would mean for constraints on the dark energy, leaving the propagation of such uncertainties through to the cosmological parameters for future work. However, it is important to note that we have considered a level of uncertainty in the photometric redshift errors that is significant for constraints on the dark energy.  The differences from the Gaussian case underscore the importance of considering realistic sources of uncertainty (although the two variations presented here are somewhat simplified examples and not intended to be taken as a full exploration of such uncertainties). We note that we have parameterized uncertainties of similar order to those considered in \citet{Ma:2005rc}, but more generally and with fewer parameters (15 vs. 40).

\section{Discussion and Conclusions}
\label{sec:conclusions}

With the potential to significantly improve our knowledge of the expansion history of the universe, future observational programs are necessary for making theoretical progress on the puzzle of dark energy. However systematic errors have the potential to wreak havoc on these results.  In particular, photometric redshift methods, essential to many future experimental programs, rely on modeling assumptions that can cause both biases and catastrophic errors in the redshift estimates.  In order for future observations to deliver on their promise, these errors must be reduced or well-understood.

In this work we have presented a method of exploring and parameterizing modeling uncertainties associated with photometric redshift estimates. We consider two sources of modeling uncertainty -- the SED template set and magnitude priors -- and we show that both cause errors in the photometrically estimated $dN/dz$.  Though both cases utilize simplified models to probe their impact, the results indicate that template selection effects are dominant to those of the magnitude priors (mean subtracted data, $d_\alpha = dN/dz_\alpha - \langle dN/dz \rangle$, are nearly an order of magnitude larger for the template variations).  This result comes with the caveat that the underlying parameterization of the prior distribution was perfectly known for the template variations and we only consider the effects of uncertainties in prior parameters (differences could be larger in the case where the functional form of the distribution is incorrect or training set is biased in some way.) Additionally, we vary the template set and the priors independently, although the effects of these variations may be correlated. A more sophisticated model would enable us to examine such modeling uncertainties jointly.

We have shown that in both cases, the variations in the $dN/dz$ can be characterized by a principal component analysis, which selects out the directions of maximum variance.  Using the principal components as the new basis, the $dN/dz$ can be parameterized by the weights, $b_i$, where the $dN/dz_i$ for each collection is constructed according to Eq. \ref{eq:reconstruct}.  The uncertainties in the model, can then be described in terms of a covariance matrix for the $b_i$ and can be used in a likelihood analysis for weak lensing or other cosmological observations involving photometric redshifts.

PCA allows for dimensionality reduction (here we find that two or three modes are sufficient to account for 80\% of the variation in the data).  We note that the precision of the reconstruction can be easily ``tuned" by retention of a larger or smaller number of modes.  The exact number of modes needed for a given level of fidelity to the original data will depend on the exact data set and modeling uncertainties considered.  However it is a simple matter to choose a desired level of variation captured by the parameterization: $\lambda_{\alpha}/\sum_{i}\lambda_{i}$, where $\lambda_i$ is the eigenvalue associated with a given mode.  For the case of five redshift bins (or collections) and three principal components, the uncertainty in the photometric redshifts can be captured by $15$ numbers (a computationally tractable result).  Moreover, this parameterization is completely general and does not resort to ad hoc forms such as Gaussians. Thus we present a way of parameterizing realistic modeling uncertainties and propagating these uncertainties through to the cosmological parameters.

More generally, this method of estimating \phz uncertainty and propagating to cosmological parameters could be applied to the more realistic observational case where two or more probes of geometry and structure formation (such as cosmic shear tomography and baryon acoustic oscillations) are combined. This cross-calibration removes degeneracies, significantly reducing sensitivity to systematic error. Finally, there are statistical calibrations that can be done in regions of the sky where there are spectroscopic samples, and this information can be imported to the method we have developed. In such cases the mean redshift of the galaxy population in every collection may be calibrated by cross-correlating that sample with a bright spectroscopic sample in angle and redshift \citep{Matthews:2010an}. Leveraging the new deep-wide spectroscopic surveys, one could go beyond a calibration of the mean redshift of that sample to knowledge of the distribution $P_{\lambda t} (z)$ of galaxies in that collection as a fuction of type. This information could then be incorporated into the prior.


\begin{acknowledgements}We would like to thank Andreas Albrecht, Jim Bosch, and Ami Choi for useful discussions, and Perry Gee for computing support.  This work was supported by the National Science Foundation grant 0709498. JAT and SS were supported by the Research Corporation. HZ is supported by the Bairen program from the Chinese Academy of Sciences, the National Basic Research Program of China grant No.~2010CB833000, and the National Science Foundation of China grant No.~11033005.

\end{acknowledgements}

\appendix
\section{Estimator Bias}
\label{sec:bias}


Here we show that the simple estimator Equation (\ref{eq:est}) is unbiased
if the prior and likelihood are both correct.
Assuming an underlying redshift distribution $dN/dz$, we get a posterior
distribution for the $i$th galaxy
\be
\label{eqn:posterior}
P_i(z)  \equiv P(z|f_{\lambda, i}^d) =
P(f_{\lambda, i}^d | z) \frac{dN}{dz} A(f_{\lambda, i}^d ),
\ee
where $A(f_{\lambda, i}^d )$ is a normalization factor so that
\be
\int dz P_i(z) = 1.
\ee
For simplicity we have suppressed $t$ \& $m_1$, which can be integrated out
via Equation (\ref{eq:int_tm}).
In what follows, we also drop the subscript $\lambda$ and superscript $d$
of $f$ for convenience.
We group galaxies that meet certain criteria, e.g., within a volume in
color space, together and call the collection a ``redshift bin".
Redshift is only one application; the collection criteria can be more
general. In terms of symbolics, we use $f \in F$ to denote the
criteria. On the one hand, $dN/dz$ in this $F$ collection is
\be
\frac{dN_F}{dz} = \int_{f \in F}df \frac{dN}{dz}P(f|z).
\ee
On the other hand, the estimator is
\be
\frac{d\hat{N}_F}{dz}=\sum_{f_i \in F} P(z | f_i).
\ee
Calculating the expectation value for this estimator is the same as
calculating its value in the limit of very large numbers of galaxies,
assuming that the galaxies follow the redshift distribution $dN/dz$.
In this large-$N$ limit we can replace
\be
\sum_{f_i \in F} \mbox{ with }
\int_{f \in F} df \int dz_* \frac{dN}{dz_*} P(f|z_*), \nonumber
\ee
so that
\bea  \nonumber
\left\langle\frac{d\hat{N}_F}{dz}\right\rangle &=&
\int_{f \in F} df \int dz_*\frac{dN}{dz_*}A(f) P(f|z_*)P(f|z)\frac{dN}{dz}\\
&=& \int_{f \in F} df \frac{dN}{dz} P(f|z) = \frac{dN_F}{dz}.
\label{eq:equiv}
\eea
Since the above derivation does not use any specific form of the
likelihood function, and since the collection $F$ is generic,
the estimator is not biased.
In reality, the prior $dN/dz$ and the likelihood $P(f|z)$ are not
accurately known, and systematic errors in them can still cause biases in
the results.

\bibliography{photozbib5}

\end{document}